# Probable alpha and $^{14}$C cluster emission from hyper Ac nuclei

K. P. Santhosh

School of Pure and Applied Physics, Kannur University, Swami Anandatheertha Campus, Payyanur 670327, India

**Abstract.**

A systematic study on the probability for the emission of $^{4}He$ and $^{14}C$ cluster from hyper $^{207-234}_{\Lambda}Ac$ and non-strange normal $^{207-234}Ac$ nuclei are performed for the first time using our fission model, the Coulomb and proximity potential model (CPPM). The predicted half lives show that hyper $^{207-234}_{\Lambda}Ac$ nuclei are unstable against $^{4}He$ emission and $^{14}C$ emission from hyper $^{217-228}_{\Lambda}Ac$ are favorable for measurement. Our study also show that hyper $^{207-234}_{\Lambda}Ac$ are stable against hyper $^{4}_{\Lambda}He$ and $^{14}_{\Lambda}C$ emission. The role of neutron shell closure (N=126) in hyper $^{214}_{\Lambda}Fr$ daughter and role of proton/ neutron shell closure $(Z \approx 82, N = 126)$ in hyper $^{210}_{\Lambda}Bi$ daughter are also revealed. As hyper-nuclei decays to normal nuclei by mesonic/non-mesonic decay and since most of the predicted half lives for $^{4}He$ and $^{14}C$ emission from normal Ac nuclei are favourable for measurement, we presume that alpha and $^{14}C$ cluster emission from hyper Ac nuclei can be detected in laboratory in a cascade (two-step) process.

**PACS.** 23.60. +e Alpha decay - 23.70. +j Heavy particle decay - 21.80.+a Hypernuclei

email: drkpsanthosh@gmail.com

## 1. Introduction

The studies on hypernuclei have recently received lot of attention as a large number of hypernuclei are produced experimentally and ground state separation energies were determined [1-6]. The characteristic feature of the hyperon is that it is free from the Pauli Exclusion Principle, and thus it can deeply penetrate into the nuclear interior. When a $\Lambda$-hyperon replaces one of the nucleons in the nucleus, the original nuclear structure changes to a system composed by the hyperon and the core of the remaining nucleons. A hyperon may modify several properties of nuclei, such as nuclear size [7, 8], the density distribution [9], deformation properties [10, 11], the neutron/proton drip-line [12, 13, 14], and fission barrier [15, 16]. The structure of hypernuclei enables us to study hyperon-nucleon interactions, which extend our knowledge on nuclear force toward unified understanding of baryon-baryon interactions.

The lifetime of hyperons bound in hypernuclei is affected by the nuclear environment. The decay of free $\Lambda$-hyperon is purely mesonic $\Lambda \rightarrow \mathrm{N} + \pi$, but in heavy hyper nuclei mesonic decay is negligible and the total decay width is due to the nonmesonic decay channels. Ohm et al [17] have shown that lifetime of the hyperon in heavy hypernuclei to be roughly of the same magnitude as for the free $\Lambda$-hyperon decay and also experimentally shown that (p,K) reaction is an effective method to produce heavy hypernuclei with large cross sections (~150mb) even at the subthreshold bombarding energy of $T_P$ = 51.5 GeV.

Production of hyper nuclei [18] can be achieved through "strangeness-exchange reaction" which requires the injection of π+ or $K^-$ beams on fixed targets (see [19] and references therein) and also electron beams on fixed targets [20, 21]. When a $K^-$ stops inside a nucleus, a neutron is replaced by a Λ hyperon with the emission of a pion. By precisely studying momentum of the outgoing pions both the binding energy and the formation probability [22] of the hyper-nuclei can be measured.

In the present paper we have made an attempt for the first time to study how hyper nuclei behave against alpha and heavy cluster emission, by studying/comparing the tunneling probability/half life of alpha and heavy cluster emission in both hypernuclei and non-strange normal nuclei. We would like to point out that many of the normal Ac isotopes are alpha emitters [23] and $^{14}$C clusters are observed from $^{223}$Ac and $^{225}$Ac isotopes [24]. Our recent study [25] shows that 14 isotopes of Ac with mass in the region A = 216-229 are favourable candidates for $^{14}$C cluster emission and this is the reason for choosing Ac isotopes for the present study.

Alpha decay was first interpreted as quantum mechanical tunneling through the potential barrier by Gamow [26] and independently by Gurney and Condon [27] in 1928. There are many effective theoretical approaches that have been used to describe alpha decay, such as Generalized Liquid Drop Model (GLDM) [28], Generalized Density dependent Cluster Model (GDDCM) [29], Unified Model for Alpha Decay and Alpha Capture (UMADAC) [30], and Coulomb and Proximity Potential Model (CPPM) [31], and all of them have been successful in reproducing the experimental data. Cluster radioactivity, the emission of particle heavier than alpha particle was first predicted by Sandulescu et al. [32] in 1980 and such decays were first observed experimentally by Rose and Jones [33] in 1984 in the radioactive decay of $^{223}Ra$ by the emission of $^{14}C$. The cluster decay process has been studied extensively using different theoretical models with different realistic nuclear interaction potentials [31, 34-41].

Using the Coulomb and Proximity Potential Model [31, 37] we have studied $^{4}He$ and $^{14}C$ clusters emission from hyper $^{207-234}_{\Lambda}Ac$ and non-strange $^{207-234}Ac$ nuclei to find the most promising hyper nuclei which are most favourable for emission. We have also studied the possibility for the emission of hyper $^{4}_{\Lambda}He$ and $^{14}_{\Lambda}C$ cluster from these parent nuclei. The Coulomb and Proximity Potential Model [31, 37] have been successful in studying alpha and cluster radioactivity in various mass regions of the nuclear chart. In this model the interacting barrier for the post scission region is taken as the sum of Coulomb and proximity potential and for the overlap region a simple power law interpolation is used.

The formalism of the Coulomb and proximity potential model (CPPM) is presented in Sec. 2. The result and discussion on the decay of hyper ${}^{207-234}_{\Lambda}Ac$ and non-strange ${}^{207-234}Ac$ nuclei are given in Sec. 3, and, in Sec. 4, we summarize the entire work.

## 2. The Coulomb and Proximity Potential Model (CPPM)

In the Coulomb and proximity potential model (CPPM), the potential energy barrier is taken as the sum of Coulomb potential, proximity potential and centrifugal potential for the touching configuration and for the separated fragments. For the pre-scission (overlap) region, simple power law interpolation as done by Shi and Swiatecki [41] is used. The inclusion of proximity potential reduces the height of the potential barrier, which closely agrees with the experimental result. The proximity potential was first used by Shi and Swiatecki [41] in an empirical manner and has been quite extensively used by Gupta et al., [40] in the Preformed Cluster Model (PCM). R K Puri et al., [42, 43] has been using different versions of proximity potential for studying fusion cross section of different target-projectile combinations. In our model contribution of both internal and external part of the barrier is considered for the penetrability calculation. In present model assault frequency, $\nu$ is calculated for each parent-cluster combination which is associated with vibration energy. But Shi and Swiatecki [44] get $\nu$ empirically, unrealistic values $10^{22}$ for even-A parents and $10^{20}$ for odd-A parents.

The interacting potential barrier for two spherical nuclei is given by

$$V = \frac{Z_1 Z_2 e^2}{r} + V_p(z) + \frac{\hbar^2 \ell(\ell+1)}{2\mu r^2} \quad , \text{for } z > 0 \tag{1}$$

Here $Z_1$ and $Z_2$ are the atomic numbers of the daughter and emitted cluster, 'z' is the distance between the near surfaces of the fragments, 'r' is the distance between fragment centres, $\ell$ represents the angular momentum, $\mu$ the reduced mass, $V_P$ is the proximity potential given by Blocki *et al.*, [45] as

$$V_p(z) = 4\pi\gamma b \left[ \frac{C_1 C_2}{(C_1 + C_2)} \right] \Phi\left( \frac{z}{b} \right) \tag{2}$$

With the nuclear surface tension coefficient,

$$\gamma = 0.9517 \, [1 - 1.7826 \, (N - Z)^2 / A^2] \quad \text{MeV/fm}^2 \tag{3}$$

where N, Z and A represent the neutron, proton and mass number of the parent, $\Phi$ represents the universal proximity potential [46] given as

$$\Phi(\varepsilon) = -4.41 e^{-\varepsilon / 0.7176} \quad , \text{for } \varepsilon \geq 1.9475 \tag{4}$$

$$\Phi(\varepsilon) = -1.7817 + 0.9270\varepsilon + 0.0169\varepsilon^2 - 0.05148\varepsilon^3 \quad , \text{for } 0 \leq \varepsilon \leq 1.9475 \tag{5}$$

with $\varepsilon = z/b$, where the width (diffuseness) of the nuclear surface $b \approx 1$ and Süsmann central radii $C_i$ of the fragments related to sharp radii $R_i$ is

$$C_i = R_i - \left(\frac{b^2}{R_i}\right) \tag{6}$$

For $R_i$ we use the semi empirical formula in terms of mass number $A_i$ as [45]

$$R_i = 1.28 A_i^{1/3} - 0.76 + 0.8 A_i^{-1/3} \tag{7}$$

The potential for the internal part (overlap region) of the barrier is given as

$$V = a_0 \left(L - L_0\right)^n \qquad \text{for } z < 0 \tag{8}$$

where $L = z + 2C_1 + 2C_2$ and $L_0 = 2C$, the diameter of the parent nuclei. The constants $a_0$ and n are determined by the smooth matching of the two potentials at the touching point.

Using one dimensional WKB approximation, the barrier penetrability P is given as

$$\mathrm{P} = \exp\left\{-\frac{2}{\hbar}\int_a^b \sqrt{2\mu(V-Q)}\,dz\right\} \tag{9}$$

Here the mass parameter is replaced by $\mu = mA_1A_2/A$, where m is the nucleon mass and $A_1$, $A_2$ are the mass numbers of daughter and emitted cluster respectively. The turning points "a" and "b" are determined from the equation, $V(a) = V(b) = Q$. The above integral can be evaluated numerically or analytically, and the half life time is given by

$$T_{1/2} = \left(\frac{\ln 2}{\lambda}\right) = \left(\frac{\ln 2}{\nu P}\right) \tag{10}$$

where, $\nu = \left(\frac{\omega}{2\pi}\right) = \left(\frac{2E_v}{h}\right)$ represent the number of assaults on the barrier per second and λ the decay constant. $E_v$, the empirical vibration energy is given as [47]

$$E_v = Q\left\{0.056 + 0.039\exp\left[\frac{(4 - A_2)}{2.5}\right]\right\} \quad , \quad \text{for } A_2 \geq 4 \tag{11}$$

In the classical method, the $\alpha$ particle is assumed to move back and forth in the nucleus and the usual way of determining the assault frequency is through the expression given by $\nu = velocity/(2R)$, where *R* is the radius of the parent nuclei. But the alpha particle has wave properties; therefore a quantum mechanical treatment is more accurate. Thus, assuming that the alpha particle vibrates in a harmonic oscillator potential with a frequency *ω*, which depends on the vibration energy $E_v$, we can identify this frequency as the assault frequency *ν* given in eqns. (10)-(11).

### 3. Results and discussion

The half lives for the emission of $^4He$ and $^{14}C$ clusters from hyper $^{207-234}_{\Lambda}Ac$ and non-strange $^{207-234}Ac$ nuclei have been calculated using the Coulomb and proximity potential model (CPPM). The decay energy of the reaction is given as

$$Q = \Delta M - (\Delta M_1 + \Delta M_2) > 0 \tag{12}$$

Here $\Delta M$ ,$\Delta M_1, \Delta M_2$ are the mass excess of the parent, daughter and cluster respectively.

A hyper nucleus can be considered as the core of a normal nucleus plus the hyperons. The binding energy of the hyper nucleus can be written as

$$B(A,Z)_{hyper} = B(A-1,Z)_{core} + S_\Lambda \quad (13)$$

where $B(A,Z)_{hyper}$ is the binding energy of a hyper nucleus, $B(A-1,Z)_{core}$ is the binding energy of its non-strange core nucleus and $S_\Lambda$ is the $\Lambda$-hyperon separation energy. For computing Q values, experimental $\Lambda$-hyperon separation energies are taken from Ref [1-6] and binding energies from latest mass tables of Wang et al [48]. For those nuclei for which experimental $\Lambda$-hyperon separation energies are not available, binding energy formula reported by Samanta et al [49] can be used to calculate hyperon separation energy. But to get better accuracy we used the formula which is obtained by least square regression to the experimental data given as

$$S_\Lambda = 0.036A + 18.90 \qquad \text{for } A \geq 50 \quad (14)$$

In Table 1 the computed Q values and logarithm of half lives for the decay of $^4He$ from hyper $^{207-234}_{\Lambda}Ac$ and non-strange normal $^{207-234}Ac$ nuclei are displayed. In the case of hypernuclei the half life values are computed taking into account of the changes in the decay Q value due to a $\Lambda$-particle, and the corresponding $\log_{10}(T_{1/2})$ values are given in Table 1 as Cal. 1. To account for the changes in the potential due to a $\Lambda$-particle we have included the potential, $V_\Lambda$ between the non-strange normal fragment and the fragment that contains lambda particle, in the expression for the interacting potential (eqn. 1). The potential, $V_\Lambda$ between the non-strange and strange fragments is given by

$$V_\Lambda = \int \rho_\Lambda(r_1) V_{\Lambda N}(r_1 - r) d^3 r_1 \qquad (15)$$

where $\rho_\Lambda(r_1)$ is the density distribution of lambda particle. The density distribution of lambda particle is taken from Ref. [7, 8] and has the form

$$\rho_\Lambda(r) = \left(\pi b_\Lambda^2\right)^{-3/2} e^{-r^2/b_\Lambda^2} \qquad (16)$$

Here $b_\Lambda = \sqrt{(4M_N + M_\Lambda)/4M_\Lambda} b_\alpha$, where $M_N$ and $M_\Lambda$ are the mass of the nucleon and $\Lambda$ particle respectively, and $b_\alpha = 1.358 fm$. The lambda-nucleon force is short range and the strength of lambda-nucleus potential $V_{\Lambda N}$ is smaller than the nucleon-nucleus potential, we have taken the lambda-nucleus potential, $V_{\Lambda N}$ from Ref [50] and given as,

$$V_{\Lambda N} = \frac{-V_0}{1 + \exp[(r - c)/a]} \qquad (17)$$

Here the constant $V_0 = 27.4 MeV$, $a = 0.6 fm$ and $c = 1.08 A^{1/3}$. By considering the changes in both Q value and potential due to a lambda particle we have recalculated the logarithm of half life values and are given in Table 1 as Cal. 2. The predicted half lives ($T_{1/2} < 10^{30}$s) show that the isotopes hyper ${}^{207-234}_{\Lambda}Ac$ nuclei are unstable against ${}^{4}He$ emission. For hyper ${}^{4}_{\Lambda}He$ emission from hyper ${}^{207-234}_{\Lambda}Ac$ nuclei the computed Q values are found to be negative which shows that these nuclei are stable against hyper ${}^{4}_{\Lambda}He$ emission. Fig 1 and Fig 2 represent the plot connecting Q value and $\log_{10}(T_{1/2})$ versus neutron number of parent nuclei for the alpha decay of hyper ${}^{208-234}_{\Lambda}Ac$ and non-strange ${}^{207-233}Ac$ nuclei respectively. The peak in Q value for alpha decay from ${}^{217}Ac$ and hyper ${}^{218}_{\Lambda}Ac$ nuclei denote the role of neutron shell closure at N=126 in daughter ${}^{213}Fr$ and ${}^{214}_{\Lambda}Fr$ daughter respectively. Also the dip (minimum) in $\log_{10}(T_{1/2})$ for alpha decay from ${}^{217}Ac$ and hyper ${}^{218}_{\Lambda}Ac$ nuclei denote the role of neutron shell closure at N=126 in the daughter nuclei.

From Fig 1 and Fig 2 it is clear that both Q value curve and alpha half lives curve of non-strange normal Ac nuclei overlap with that of hyper Ac nuclei. i.e. When the decay half lives of non-strange normal Ac nuclei are compared with that of corresponding hyper Ac nuclei with same neutron number (for e.g. half lives of $^{217}Ac$ and $^{218}_{\Lambda}Ac$ nuclei) it can be seen that the half lives do not differ much.

In Table 2 the computed Q values and $\log_{10}(T_{1/2})$ values for the $^{14}C$ cluster emission from hyper $^{207-224}_{\Lambda}Ac$ and non-strange $^{207-234}Ac$ are given. Here also Cal. 1 denotes the $\log_{10}(T_{1/2})$ values taking into account of the changes in the decay Q value due to a $\Lambda$-particle and Cal. 2 denotes the $\log_{10}(T_{1/2})$ values taking into account of the change in both Q value and potential due to a $\Lambda$-particle.

Fig 3 and Fig 4 represent the plot connecting Q value and $\log_{10}(T_{1/2})$ versus neutron number of parent nuclei for the $^{14}C$ cluster decay of hyper $^{207-233}_{\Lambda}Ac$ and non-strange $^{208-234}Ac$ nuclei respectively. The peak in Q value and dip in $\log_{10}(T_{1/2})$ for $^{14}C$ cluster decay from $^{223}Ac$ stress the role of near doubly magic $^{209}Bi$ daughter $(Z \approx 82, N = 126)$ in cluster decay. Also the peak in Q value and dip in $\log_{10}(T_{1/2})$ for $^{14}C$ cluster decay from hyper $^{224}_{\Lambda}Ac$ stress the role of proton and neutron shell closure $(Z \approx 82, N = 126)$ in $^{210}_{\Lambda}Bi$ daughter.

From Fig 3 and Fig 4 it is clear that both Q value curve and cluster decay half lives curve of non-strange normal Ac nuclei almost coincide with that of hyper Ac nuclei. i.e. When the decay half lives of non-strange normal Ac nuclei are compared with that of corresponding hyper Ac nuclei with same neutron number it can be seen that the half lives do not differ much.

We have compared the computed logarithm of alpha half lives for normal Ac isotopes with experimental data [23] given in the last column of Table 1. The computed values are in agreement with experimental data and the standard deviation of logarithm of computed alpha half is found to be 0.72. The standard deviation is computed using the relation

$$\sigma = \sqrt{\frac{1}{(n-1)}\sum_{i=1}^{n}\left[\log\left(T_{1/2}^{cal.}\right)-\log\left(T_{1/2}^{\exp.}\right)\right]^2} \qquad (18)$$

We have also compared the half lives for $^{14}C$ emission from normal Ac isotopes with experimental values, From Table 2 it is clear that the computed $\log_{10}(T_{1/2})$ values for $^{14}C$ emission from normal $^{223}Ac$ and $^{225}Ac$ are in good agreement with experimental values [24].

Fig 5(a) and 5(b) represents the plot connecting Q value and $\log_{10}(T_{1/2})$ versus mass number of parent nuclei respectively for the ${}^{14}_{\Lambda}C$ cluster decay of hyper ${}^{207-224}_{\Lambda}Ac$ nuclei. The peak in Q value and dip (minimum) in $\log_{10}(T_{1/2})$ for ${}^{14}_{\Lambda}C$ cluster decay from hyper ${}^{223}_{\Lambda}Ac$ indicate the role of near doubly magic $^{209}Bi$ daughter $(Z \approx 82, N = 126)$ in the cluster decay. The computed half life for hyper ${}^{14}_{\Lambda}C$ emission from hyper ${}^{207-224}_{\Lambda}Ac$ nuclei is found to have large values, $T_{1/2} > 10^{66}$s which shows that hyper ${}^{207-224}_{\Lambda}Ac$ nuclei are stable against hyper ${}^{14}_{\Lambda}C$ emission. The predicted ${}^{14}C$ decay half life for ${}^{217-228}_{\Lambda}Ac$ are well within the present upper limit for measurement $(T_{1/2} < 10^{30} s)$ and these decays are favourable for measurement.

## 4. Summary

The Coulomb and Proximity Potential Model (CPPM) have been used to find possibilities of alpha and ${}^{14}C$ cluster emission from hyper ${}^{207-234}_{\Lambda}Ac$ nuclei. The Q values are calculated using the recent mass tables of Wang et al [48]. The predicted half lives show that hyper ${}^{207-234}_{\Lambda}Ac$ nuclei are unstable against ${}^{4}He$ emission and ${}^{14}C$ emission from ${}^{217-228}_{\Lambda}Ac$ are favorable for measurement. Our study also show that hyper ${}^{207-234}_{\Lambda}Ac$ are stable against hyper ${}^{4}_{\Lambda}He$ and ${}^{14}_{\Lambda}C$ emission. The role of neutron shell closure (N=126) in ${}^{214}_{\Lambda}Fr$ daughter and role of proton/ neutron shell closure $(Z \approx 82, N = 126)$ in ${}^{210}_{\Lambda}Bi$ daughter are also revealed. The computed alpha and $^{14}$C cluster half lives from normal Ac nuclei are compared with corresponding experimental data and are found to be in good agreement. As hyper-nuclei decays to normal nuclei by mesonic/non-mesonic decay and since most of the predicted half lives for ${}^{4}He$ and ${}^{14}C$ emission from normal Ac nuclei are favourable for measurement, we presume that alpha and $^{14}C$ cluster emission from hyper Ac nuclei can be detected in laboratory in a cascade (two-step) process.

**Table 1.** Computed Q value and logarithm of half lives for the decay of ${}^{4}He$ from hyper ${}^{207-234}_{\Lambda}Ac$ and non-strange ${}^{207-234}Ac$ nuclei. $T_{1/2}$ is in second.

| Decay | Q Value (MeV) | $\log_{10}(T_{1/2})$ | | Decay | Q Value (MeV) | $\log_{10}(T_{1/2})$ | |
|---|---|---|---|---|---|---|---|
| | | Cal. 1 | Cal. 2 | | | Present | Expt.[23] |
| ${}^{207}_{\Lambda}Ac \rightarrow {}^{4}He + {}^{203}_{\Lambda}Fr$ | 7.782 | -0.758 | -1.158 | ${}^{207}Ac \rightarrow {}^{4}He + {}^{203}Fr$ | 7.849 | -0.978 | -1.509 |
| ${}^{208}_{\Lambda}Ac \rightarrow {}^{4}He + {}^{204}_{\Lambda}Fr$ | 7.701 | -0.496 | -0.896 | ${}^{208}Ac \rightarrow {}^{4}He + {}^{204}Fr$ | 7.728 | -0.577 | -1.009 |
| ${}^{209}_{\Lambda}Ac \rightarrow {}^{4}He + {}^{205}_{\Lambda}Fr$ | 7.582 | -0.089 | -0.489 | ${}^{209}Ac \rightarrow {}^{4}He + {}^{205}Fr$ | 7.725 | -0.588 | -1.032 |
| ${}^{210}_{\Lambda}Ac \rightarrow {}^{4}He + {}^{211}_{\Lambda}Fr$ | 7.586 | -0.124 | -0.523 | ${}^{210}Ac \rightarrow {}^{4}He + {}^{211}Fr$ | 7.607 | -0.187 | -0.415 |
| ${}^{211}_{\Lambda}Ac \rightarrow {}^{4}He + {}^{207}_{\Lambda}Fr$ | 7.462 | 0.307 | -0.091 | ${}^{211}Ac \rightarrow {}^{4}He + {}^{207}Fr$ | 7.619 | -0.251 | -0.672 |
| ${}^{212}_{\Lambda}Ac \rightarrow {}^{4}He + {}^{208}_{\Lambda}Fr$ | 7.478 | 0.227 | -0.170 | ${}^{212}Ac \rightarrow {}^{4}He + {}^{208}Fr$ | 7.521 | 0.085 | 0.208 |
| ${}^{213}_{\Lambda}Ac \rightarrow {}^{4}He + {}^{209}_{\Lambda}Fr$ | 7.374 | 0.597 | 0.199 | ${}^{213}Ac \rightarrow {}^{4}He + {}^{209}Fr$ | 7.503 | 0.132 | -0.136 |
| ${}^{214}_{\Lambda}Ac \rightarrow {}^{4}He + {}^{210}_{\Lambda}Fr$ | 7.355 | 0.650 | 0.253 | ${}^{214}Ac \rightarrow {}^{4}He + {}^{210}Fr$ | 7.353 | 0.674 | 0.964 |
| ${}^{215}_{\Lambda}Ac \rightarrow {}^{4}He + {}^{211}_{\Lambda}Fr$ | 7.209 | 1.194 | 0.798 | ${}^{215}Ac \rightarrow {}^{4}He + {}^{211}Fr$ | 7.746 | -0.782 | -0.770 |
| ${}^{216}_{\Lambda}Ac \rightarrow {}^{4}He + {}^{212}_{\Lambda}Fr$ | 7.603 | -0.305 | -0.698 | ${}^{216}Ac \rightarrow {}^{4}He + {}^{212}Fr$ | 9.236 | -5.394 | -3.357 |
| ${}^{217}_{\Lambda}Ac \rightarrow {}^{4}He + {}^{213}_{\Lambda}Fr$ | 9.090 | -5.028 | -5.427 | ${}^{217}Ac \rightarrow {}^{4}He + {}^{213}Fr$ | 9.832 | -6.950 | -7.161 |
| ${}^{218}_{\Lambda}Ac \rightarrow {}^{4}He + {}^{214}_{\Lambda}Fr$ | 9.687 | -6.623 | -7.022 | ${}^{218}Ac \rightarrow {}^{4}He + {}^{214}Fr$ | 9.373 | -5.797 | -5.967 |
| ${}^{219}_{\Lambda}Ac \rightarrow {}^{4}He + {}^{215}_{\Lambda}Fr$ | 9.233 | -5.456 | -5.846 | ${}^{219}Ac \rightarrow {}^{4}He + {}^{215}Fr$ | 8.827 | -4.303 | -4.928 |
| ${}^{220}_{\Lambda}Ac \rightarrow {}^{4}He + {}^{216}_{\Lambda}Fr$ | 8.682 | -3.913 | -4.297 | ${}^{220}Ac \rightarrow {}^{4}He + {}^{216}Fr$ | 8.348 | -2.874 | -1.579 |
| ${}^{221}_{\Lambda}Ac \rightarrow {}^{4}He + {}^{217}_{\Lambda}Fr$ | 8.204 | -2.450 | -2.834 | ${}^{221}Ac \rightarrow {}^{4}He + {}^{217}Fr$ | 7.780 | -1.008 | -1.284 |

| Decay | Q Value (MeV) | $\log_{10}(T_{1/2})$ | | Decay | Q Value (MeV) | $\log_{10}(T_{1/2})$ | |
|---|---|---|---|---|---|---|---|
| | | Cal. 1 | Cal. 2 | | | Present | Expt.[23] |
| ${}^{222}_{\Lambda}Ac \rightarrow {}^{4}He + {}^{218}_{\Lambda}Fr$ | 7.638 | -0.542 | -0.930 | ${}^{222}Ac \rightarrow {}^{4}He + {}^{218}Fr$ | 7.138 | 1.366 | 0.703 |
| ${}^{223}_{\Lambda}Ac \rightarrow {}^{4}He + {}^{219}_{\Lambda}Fr$ | 6.993 | 1.913 | 1.523 | ${}^{223}Ac \rightarrow {}^{4}He + {}^{219}Fr$ | 6.783 | 2.812 | 2.105 |
| ${}^{224}_{\Lambda}Ac \rightarrow {}^{4}He + {}^{220}_{\Lambda}Fr$ | 6.640 | 3.399 | 3.010 | ${}^{224}Ac \rightarrow {}^{4}He + {}^{220}Fr$ | 6.327 | 4.850 | 5.027 |
| ${}^{225}_{\Lambda}Ac \rightarrow {}^{4}He + {}^{221}_{\Lambda}Fr$ | 6.183 | 5.509 | 5.123 | ${}^{225}Ac \rightarrow {}^{4}He + {}^{221}Fr$ | 5.935 | 6.783 | 5.937 |
| ${}^{226}_{\Lambda}Ac \rightarrow {}^{4}He + {}^{222}_{\Lambda}Fr$ | 5.791 | 7.512 | 7.129 | ${}^{226}Ac \rightarrow {}^{4}He + {}^{222}Fr$ | 5.535 | 8.963 | 9.246 |
| ${}^{227}_{\Lambda}Ac \rightarrow {}^{4}He + {}^{223}_{\Lambda}Fr$ | 5.391 | 9.777 | 9.399 | ${}^{227}Ac \rightarrow {}^{4}He + {}^{223}Fr$ | 5.042 | 12.002 | 10.697 |
| ${}^{228}_{\Lambda}Ac \rightarrow {}^{4}He + {}^{224}_{\Lambda}Fr$ | 4.898 | 12.944 | 12.571 | ${}^{228}Ac \rightarrow {}^{4}He + {}^{224}Fr$ | 4.677 | 14.555 | |
| ${}^{229}_{\Lambda}Ac \rightarrow {}^{4}He + {}^{225}_{\Lambda}Fr$ | 4.532 | 15.614 | 15.247 | ${}^{229}Ac \rightarrow {}^{4}He + {}^{225}Fr$ | 4.452 | 16.269 | |
| ${}^{230}_{\Lambda}Ac \rightarrow {}^{4}He + {}^{226}_{\Lambda}Fr$ | 4.308 | 17.412 | 17.048 | ${}^{230}Ac \rightarrow {}^{4}He + {}^{226}Fr$ | 3.872 | 21.395 | |
| ${}^{231}_{\Lambda}Ac \rightarrow {}^{4}He + {}^{227}_{\Lambda}Fr$ | 3.729 | 22.803 | 22.447 | ${}^{231}Ac \rightarrow {}^{4}He + {}^{227}Fr$ | 3.652 | 23.642 | |
| ${}^{232}_{\Lambda}Ac \rightarrow {}^{4}He + {}^{228}_{\Lambda}Fr$ | 3.507 | 25.211 | 24.859 | ${}^{232}Ac \rightarrow {}^{4}He + {}^{228}Fr$ | 3.360 | 26.964 | |
| ${}^{233}_{\Lambda}Ac \rightarrow {}^{4}He + {}^{229}_{\Lambda}Fr$ | 3.216 | 28.736 | 28.388 | ${}^{233}Ac \rightarrow {}^{4}He + {}^{229}Fr$ | 3.209 | 28.851 | |
| ${}^{234}_{\Lambda}Ac \rightarrow {}^{4}He + {}^{230}_{\Lambda}Fr$ | 3.065 | 30.752 | 30.407 | ${}^{234}Ac \rightarrow {}^{4}He + {}^{230}Fr$ | 2.905 | 33.101 | |

**Table 2.** Computed Q value and logarithm of half lives for the decay of $^{14}C$ from hyper $^{207-234}_{\Lambda}Ac$ and non-strange $^{207-234}Ac$ nuclei. $T_{1/2}$ is in second.

| Decay | Q Value (MeV) | $\log_{10}(T_{1/2})$ | | Decay | Q Value (MeV) | $\log_{10}(T_{1/2})$ | |
|---|---|---|---|---|---|---|---|
| | | Cal. 1 | Cal. 2 | | | Present | Expt.[24] |
| $^{207}_{\Lambda}Ac \rightarrow {}^{14}C + {}^{193}_{\Lambda}Bi$ | 23.466 | 39.389 | 39.535 | $^{207}Ac \rightarrow {}^{14}C + {}^{193}Bi$ | 24.003 | 37.726 | |
| $^{208}_{\Lambda}Ac \rightarrow {}^{14}C + {}^{194}_{\Lambda}Bi$ | 23.495 | 39.215 | 39.360 | $^{208}Ac \rightarrow {}^{14}C + {}^{194}Bi$ | 23.780 | 38.420 | |
| $^{209}_{\Lambda}Ac \rightarrow {}^{14}C + {}^{195}_{\Lambda}Bi$ | 23.220 | 40.121 | 40.265 | $^{209}Ac \rightarrow {}^{14}C + {}^{195}Bi$ | 23.846 | 38.125 | |
| $^{210}_{\Lambda}Ac \rightarrow {}^{14}C + {}^{196}_{\Lambda}Bi$ | 23.345 | 39.606 | 39.751 | $^{210}Ac \rightarrow {}^{14}C + {}^{196}Bi$ | 23.779 | 38.288 | |
| $^{211}_{\Lambda}Ac \rightarrow {}^{14}C + {}^{197}_{\Lambda}Bi$ | 23.274 | 39.795 | 39.938 | $^{211}Ac \rightarrow {}^{14}C + {}^{197}Bi$ | 23.867 | 37.920 | |
| $^{212}_{\Lambda}Ac \rightarrow {}^{14}C + {}^{198}_{\Lambda}Bi$ | 23.365 | 39.408 | 39.551 | $^{212}Ac \rightarrow {}^{14}C + {}^{198}Bi$ | 23.629 | 38.679 | |
| $^{213}_{\Lambda}Ac \rightarrow {}^{14}C + {}^{199}_{\Lambda}Bi$ | 23.124 | 40.204 | 40.347 | $^{213}Ac \rightarrow {}^{14}C + {}^{199}Bi$ | 23.937 | 37.553 | |
| $^{214}_{\Lambda}Ac \rightarrow {}^{14}C + {}^{200}_{\Lambda}Bi$ | 23.429 | 39.054 | 39.196 | $^{214}Ac \rightarrow {}^{14}C + {}^{200}_{\Lambda}Bi$ | 23.796 | 37.976 | |
| $^{215}_{\Lambda}Ac \rightarrow {}^{14}C + {}^{201}_{\Lambda}Bi$ | 23.293 | 39.476 | 39.618 | $^{215}Ac \rightarrow {}^{14}C + {}^{201}Bi$ | 24.426 | 35.783 | |
| $^{216}_{\Lambda}Ac \rightarrow {}^{14}C + {}^{202}_{\Lambda}Bi$ | 23.922 | 37.219 | 37.362 | $^{216}Ac \rightarrow {}^{14}C + {}^{202}Bi$ | 25.866 | 31.128 | |
| $^{217}_{\Lambda}Ac \rightarrow {}^{14}C + {}^{203}_{\Lambda}Bi$ | 25.361 | 32.430 | 32.575 | $^{217}Ac \rightarrow {}^{14}C + {}^{203}Bi$ | 27.208 | 27.100 | |
| $^{218}_{\Lambda}Ac \rightarrow {}^{14}C + {}^{204}_{\Lambda}Bi$ | 26.703 | 28.292 | 28.439 | $^{218}Ac \rightarrow {}^{14}C + {}^{204}Bi$ | 28.466 | 23.562 | |
| $^{219}_{\Lambda}Ac \rightarrow {}^{14}C + {}^{205}_{\Lambda}Bi$ | 27.964 | 24.652 | 24.802 | $^{219}Ac \rightarrow {}^{14}C + {}^{205}Bi$ | 29.614 | 20.509 | |
| $^{220}_{\Lambda}Ac \rightarrow {}^{14}C + {}^{206}_{\Lambda}Bi$ | 29.109 | 21.531 | 21.683 | $^{220}Ac \rightarrow {}^{14}C + {}^{206}Bi$ | 30.752 | 17.635 | |
| $^{221}_{\Lambda}Ac \rightarrow {}^{14}C + {}^{207}_{\Lambda}Bi$ | 30.249 | 18.588 | 18.742 | $^{221}Ac \rightarrow {}^{14}C + {}^{207}Bi$ | 31.554 | 15.678 | |

| Decay | Q Value (MeV) | $\log_{10}(T_{1/2})$ | | Decay | Q Value (MeV) | $\log_{10}(T_{1/2})$ | |
|---|---|---|---|---|---|---|---|
| | | Cal. 1 | Cal. 2 | | | Present | Expt.[24] |
| $^{222}_{\Lambda}Ac \rightarrow {}^{14}C + {}^{208}_{\Lambda}Bi$ | 31.053 | 16.581 | 16.736 | $^{222}Ac \rightarrow {}^{14}C + {}^{208}Bi$ | 32.472 | 13.526 | |
| $^{223}_{\Lambda}Ac \rightarrow {}^{14}C + {}^{209}_{\Lambda}Bi$ | 31.968 | 14.389 | 14.546 | $^{223}Ac \rightarrow {}^{14}C + {}^{209}Bi$ | 33.065 | 12.158 | 12.60 |
| $^{224}_{\Lambda}Ac \rightarrow {}^{14}C + {}^{210}_{\Lambda}Bi$ | 32.561 | 12.993 | 13.151 | $^{224}Ac \rightarrow {}^{14}C + {}^{210}Bi$ | 32.007 | 14.474 | |
| $^{225}_{\Lambda}Ac \rightarrow {}^{14}C + {}^{211}_{\Lambda}Bi$ | 31.503 | 15.361 | 15.516 | $^{225}Ac \rightarrow {}^{14}C + {}^{211}Bi$ | 30.477 | 18.037 | 17.16 |
| $^{226}_{\Lambda}Ac \rightarrow {}^{14}C + {}^{212}_{\Lambda}Bi$ | 29.973 | 19.011 | 19.162 | $^{226}Ac \rightarrow {}^{14}C + {}^{212}Bi$ | 29.407 | 20.668 | |
| $^{227}_{\Lambda}Ac \rightarrow {}^{14}C + {}^{213}_{\Lambda}Bi$ | 28.904 | 21.704 | 21.853 | $^{227}Ac \rightarrow {}^{14}C + {}^{213}Bi$ | 28.061 | 24.187 | |
| $^{228}_{\Lambda}Ac \rightarrow {}^{14}C + {}^{214}_{\Lambda}Bi$ | 27.557 | 25.316 | 25.461 | $^{228}Ac \rightarrow {}^{14}C + {}^{214}Bi$ | 27.077 | 26.903 | |
| $^{229}_{\Lambda}Ac \rightarrow {}^{14}C + {}^{215}_{\Lambda}Bi$ | 26.572 | 28.108 | 28.251 | $^{229}Ac \rightarrow {}^{14}C + {}^{215}Bi$ | 26.029 | 29.954 | |
| $^{230}_{\Lambda}Ac \rightarrow {}^{14}C + {}^{216}_{\Lambda}Bi$ | 25.526 | 31.244 | 31.385 | $^{230}Ac \rightarrow {}^{14}C + {}^{216}Bi$ | 24.944 | 33.309 | |
| $^{231}_{\Lambda}Ac \rightarrow {}^{14}C + {}^{217}_{\Lambda}Bi$ | 24.441 | 34.700 | 34.838 | $^{231}Ac \rightarrow {}^{14}C + {}^{217}Bi$ | 24.013 | 36.355 | |
| $^{232}_{\Lambda}Ac \rightarrow {}^{14}C + {}^{218}_{\Lambda}Bi$ | 23.508 | 37.845 | 37.980 | $^{232}Ac \rightarrow {}^{14}C + {}^{218}Bi$ | 22.918 | 40.171 | |
| $^{233}_{\Lambda}Ac \rightarrow {}^{14}C + {}^{219}_{\Lambda}Bi$ | 22.416 | 41.777 | 41.910 | $^{233}Ac \rightarrow {}^{14}C + {}^{219}Bi$ | 22.008 | 43.544 | |
| $^{234}_{\Lambda}Ac \rightarrow {}^{14}C + {}^{220}_{\Lambda}Bi$ | 21.401 | 45.685 | 45.816 | $^{234}Ac \rightarrow {}^{14}C + {}^{220}Bi$ | 21.001 | 47.527 | |

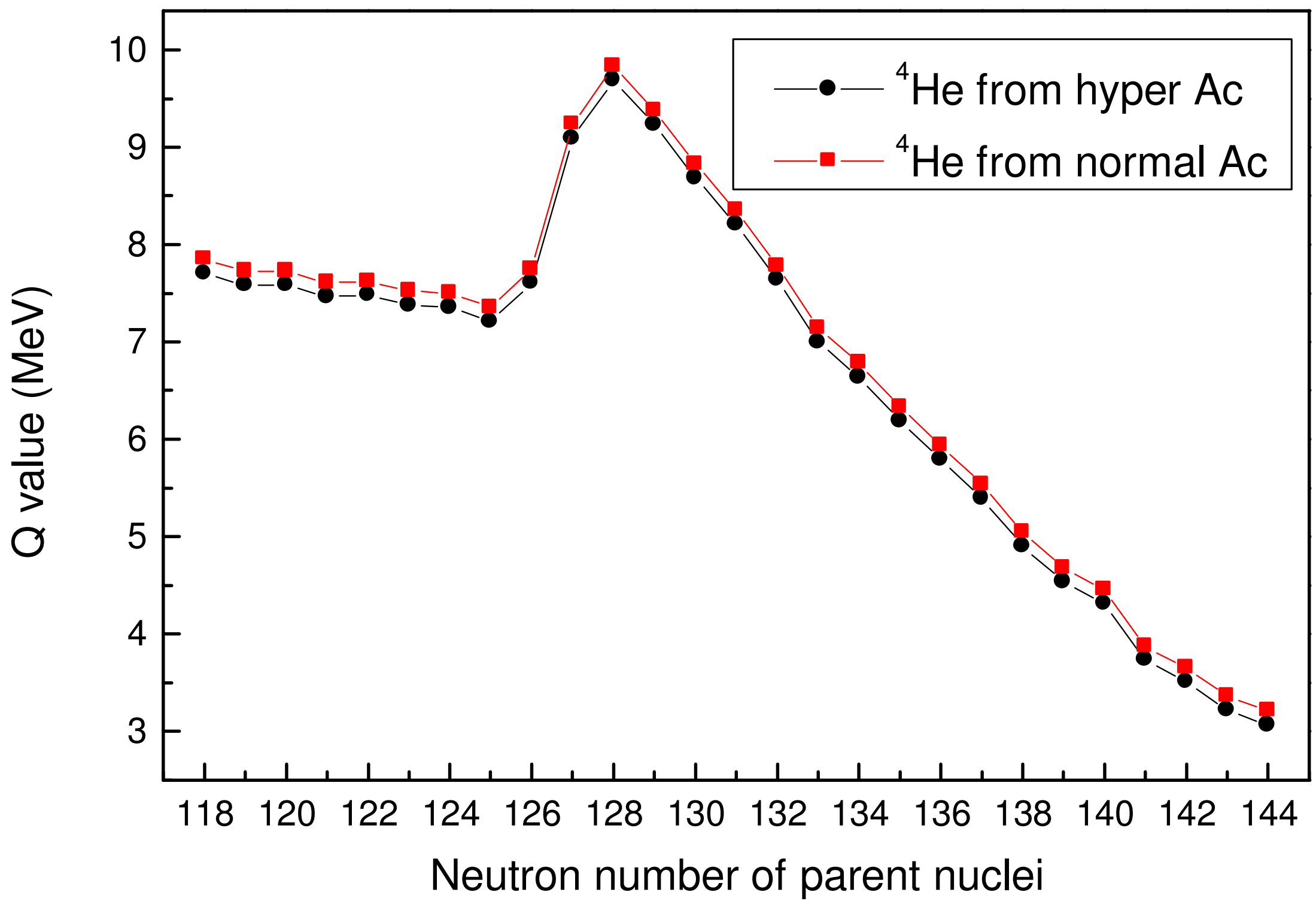

**Fig. 1.** Plot connecting Q value versus neutron number for the decay of $^{4}He$ from hyper $^{207-233}_{\Lambda}Ac$ and non-strange $^{208-234}Ac$ nuclei.

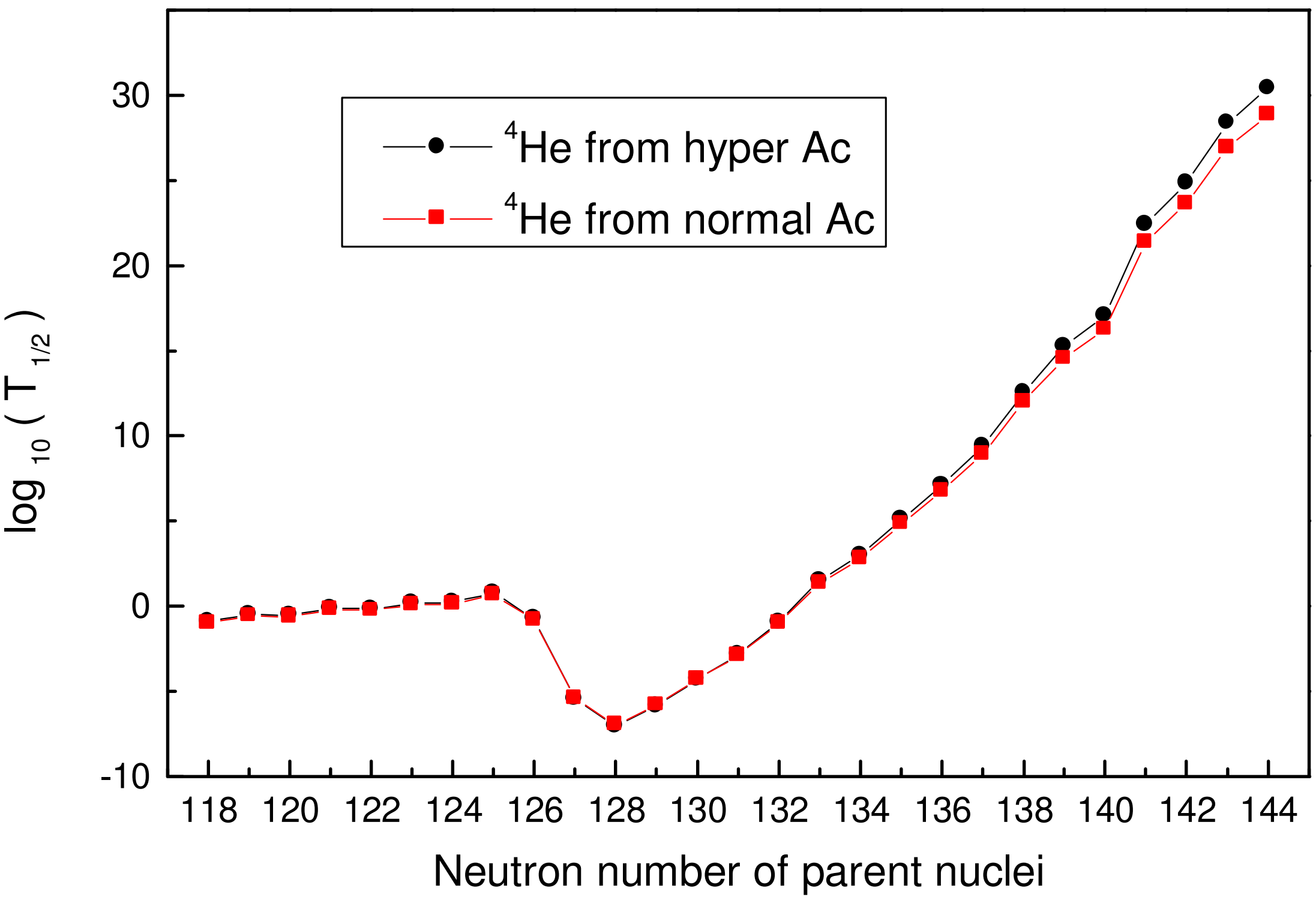

**Fig. 2.** Plot connecting $\log_{10}(T_{1/2})$ versus neutron number for the decay of $^{4}He$ from hyper $^{207-233}_{\Lambda}Ac$ and non-strange $^{208-234}Ac$ nuclei.

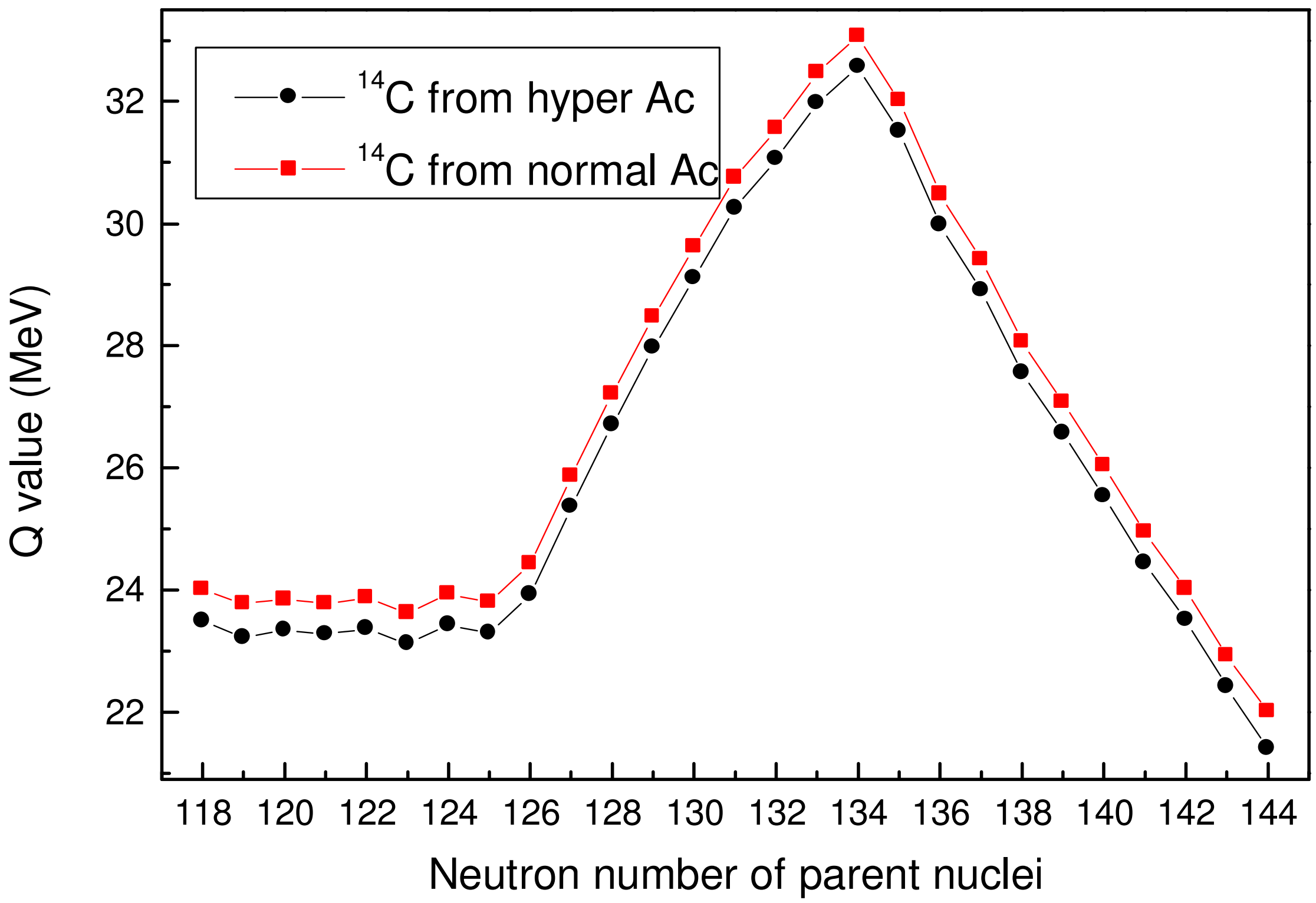

**Fig. 3.** Plot connecting Q value versus neutron number for the decay of $^{14}C$ cluster from hyper $^{207-233}_{\Lambda}Ac$ and non-strange $^{208-234}Ac$ nuclei.

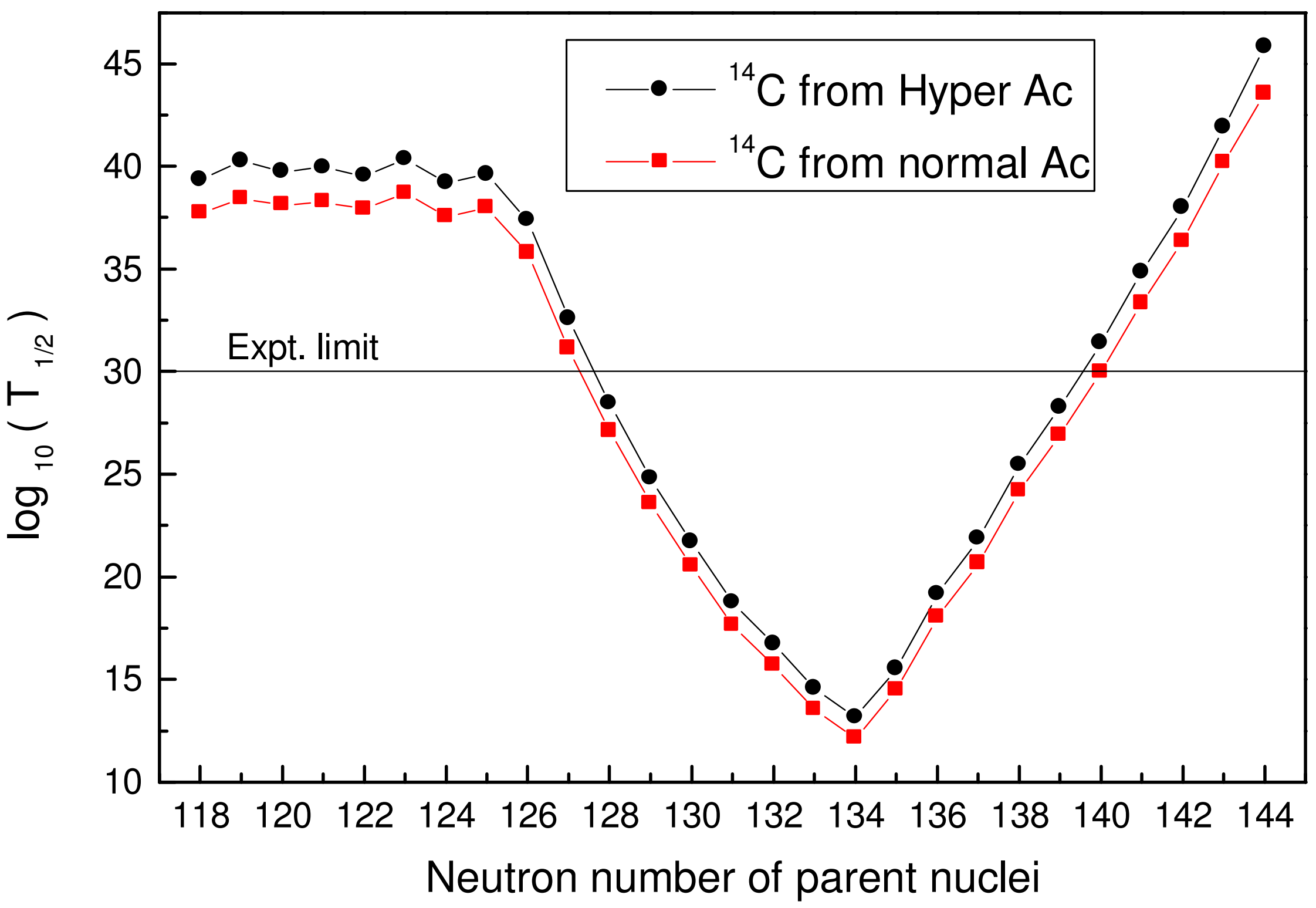

**Fig. 4.** Plot connecting $\log_{10}(T_{1/2})$ versus neutron number for the decay of $^{14}C$ cluster from hyper $^{207-233}_{\Lambda}Ac$ and non-strange $^{208-234}Ac$ nuclei.

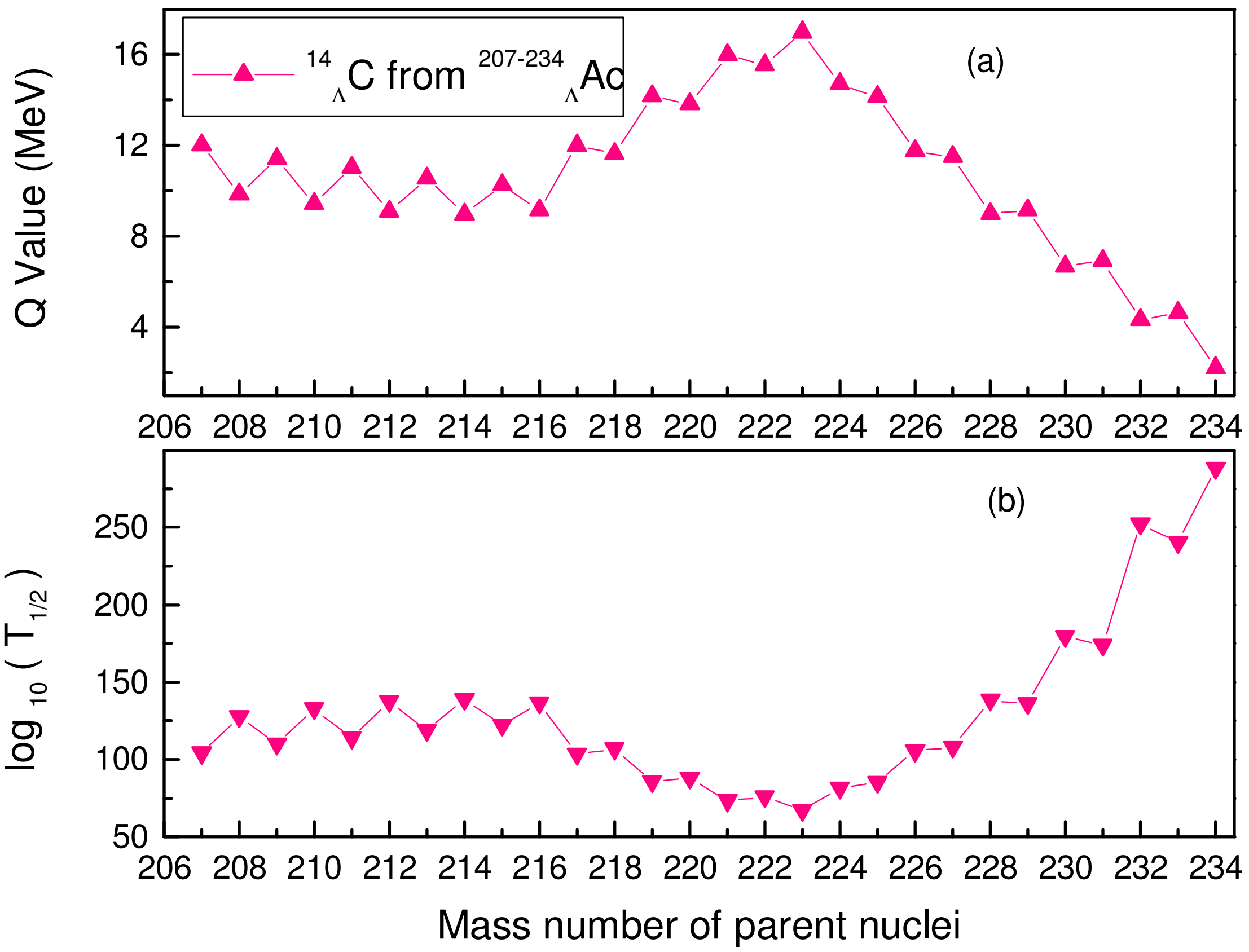

**Fig. 5.** Plot connecting (a) Q value versus mass number and (b) $\log_{10}(T_{1/2})$ versus mass number for the decay of hyper ${}^{14}_{\Lambda}C$ cluster from hyper ${}^{207-234}_{\Lambda}Ac$ nuclei.